\documentclass[a4paper,10pt]{article}
\usepackage[utf8]{inputenc}
\usepackage{amsmath,graphicx}
\usepackage[nolist,nohyperlinks]{acronym}
\usepackage{amsfonts}
\usepackage{amssymb}
\usepackage{mathtools}
\usepackage{multirow}
\usepackage[table]{xcolor}
\usepackage{tabularx}
\usepackage[textwidth=1.1\textwidth,textheight=\textheight]{geometry}
\usepackage{sectsty}
\usepackage{cleveref}
\sectionfont{\normalsize}
\subsectionfont{\normalsize}

\title{\normalsize This work was accepted and presented in: 2022 ICPR-Workshop on Artificial Intelligence for Multimedia Forensics and Disinformation Detection. Montreal, Quebec, Canada. However, due to a technical issue on the publishing companies' side, the work does not appear in the workshop proceedings. \\ \vspace{0.4cm}
\LARGE Device (In)Dependence of Deep Learning-based Image Age Approximation}
\author{\normalsize Robert Jöchl and Andreas Uhl\\\normalsize University of Salzburg, Department of Artificial Intelligence and Human Interfaces, \\\normalsize Salzburg, Austria\\
\ \normalsize \{robert.joechl, andreas.uhl\}@plus.ac.at}
\date{}

\begin{document}
\begin{acronym}
    \acro{BN}{Batch Normalization}
    \acro{CAM}{Class Activation Map}
    \acro{CNN}{Convolutional Neural Network}
    \acro{FL}{Fully Connected Layer}
    \acro{ReLU}{Rectified Linear Unit}
\end{acronym}

\maketitle

\begin{abstract}
The goal of temporal image forensic is to approximate the age of a digital image relative to images from the same device. Usually, this is based on traces left during the image acquisition pipeline. For example, several methods exist that exploit the presence of in-field sensor defects for this purpose. In addition to these `classical’ methods, there is also an approach in which a Convolutional Neural Network (CNN) is trained to approximate the image age. One advantage of a CNN is that it independently learns the age features used. This would make it possible to exploit other (different) age traces in addition to the known ones (\textit{i.e.}, in-field sensor defects). In a previous work, we have shown that the presence of strong in-field sensor defects is irrelevant for a CNN to predict the age class. Based on this observation, the question arises how device (in)dependent the learned features are. In this work, we empirically asses this by training a network on images from a single device and then apply the trained model to images from different devices. This evaluation is performed on 14 different devices, including 10 devices from the publicly available `Northumbria Temporal Image Forensics’ database. These 10 different devices are based on five different device pairs (\textit{i.e.}, with the identical camera model).
\end{abstract}
\section{Introduction}
In temporal image forensics, the main objective is to establish a chronological sequence among pieces of evidence. Since the timestamp stored in the EXIF header is easy to manipulate, this chronological order must be determined using time-dependent traces hidden in a digital image. For example, such traces are in-field sensor defects. In-field sensor defects are single pixel defects that develop in-field (\textit{i.e.,} after the manufacturing process). The characteristics of such defects are studied in multiple publications (\textit{e.g.,} \cite{chapman2016,chapman2013, leung2009B, leung2009,  leung2007, theuwissen2007, theuwissen2008}). In principle, neutrons, which are part of cosmic radiation, can damage a small part of the pixel’s silicon layer. This damage can lead to an offset independent of the incident light (\textit{i.e.,} the dynamic range of the pixel is reduced). Usually, the defect effects a single pixel only. However, because of preprocessing like demosaicing (interpolation) the defect is spread to its neighbouring pixels. Once a pixel is defective, it can no longer heal. For this reason, in-field sensor defects accumulate over time, and the age of a digital image can be approximated based on the defects detected.

A technique to approximate the age of an image based on the presence of in-field sensor defects was introduced by Fridrich and Goljan in \cite{fridrich11a}. The assumption is that a forensics analyst is provided with a set of trusted images and a second untrusted set (from the same device). The goal is to approximate the age of images from the untrusted set relative to the trusted set. For this purpose, the authors propose a maximum likelihood approach based on median filter residuals. The median filter is a denoising filter. Since in-field sensor defects appear as image noise, the median filter residuals contain the defect magnitude.

Approximating the age of a digital image can be considered a multi-class classification problem. When age approximation is based on the presence of in-field sensor defects, the classes are defined by the different defect onset times and the available trusted images. In a previous work\cite{joechl20a}, we propose to utilize traditional machine learning techniques (\textit{i.e.,} a `Naive Bayes Classifier’ and a `Support Vector Machine’) to approximate the image age based on the median filter residuals. A limitation of both approaches (\textit{i.e.,} introduced in \cite{fridrich11a} and \cite{joechl20a}) is that the defect locations must be known beforehand.

For the purpose of sensor defect detection based on regular scene images, multiple methods exist (\textit{e.g.,} \cite{chan2009, chen2012, cho2011, elyamany2017, ghosh2008, leung2009, tchendjou2020}). In \cite{Joechl21a}, we introduced a method for defect detection specifically proposed in the context of image age approximation. Another machine learning approach based on defective pixels was proposed by Ahmed et al. in \cite{ahmed21a}. In their work, the authors combine defect detection and age approximation in a single method.

In contrast to these traditional techniques, a \ac{CNN} learns the classification features used. Approximating the age of a digital image using \acp{CNN} was proposed by Ahmed et al. in \cite{ahmed20a}. The authors utilize two well-known \ac{CNN} architectures for this purpose (\textit{i.e.,} the AlexNet\cite{krizhevsky12} and GoogLeNet \cite{szegedy15a}). For a five-class classification problem, an accuracy of over 85\% was reached by the AlexNet in a transfer learning mode. The authors suggest that the learned features are not position dependent, but did not investigate this further. In \cite{Joechl21b} we systematically investigated the influence of the presence of strong in-field sensor defects on training a \ac{CNN} in the context of image age approximation. Considering the investigated `five-crop-fusion’ scenario (where five networks are trained on different fixed image patches each) the presence of a strong in-field sensor defect is irrelevant for improving the classification accuracy. For this reason, we suggested that other `age’ traces are exploited by the network. Furthermore, we evaluated if the learned features are position dependent. By showing a performance decrease when positional variance was added, we concluded that these learned features are most likely not positionally invariant. However, other properties of these learned features are not further investigated.

In this work, we empirically investigate whether these learned features are entirely device dependent. For this purpose, we train a \ac{CNN} (based on the `five-crop-fusion’ scenario defined in \cite{Joechl21b}) on images from a specific device. The trained model is then applied to approximate the age of images from different devices. In total, we perform this evaluation on 14 different devices. 10 of these 14 devices are from the publicly available `Northumbria Temporal Image Forensics’ dataset. These 10 different devices are based on five different device pairs (\textit{i.e.}, with the identical camera model). This allows us to further analyse if there are any model specific dependencies observable.

The remainder of this paper is organized as follows: the used network architecture is described in detail in section \ref{sec:Net}. In section \ref{sec:Exp}, we give an overview of the experiments conducted and the datasets used. The experimental results are stated and discussed in section \ref{sec:Results}. Finally, the key insights are summarized in the last section.

\section{Steganalysis Residual Network (SRNet)}
\label{sec:Net}
The age traces that are hidden in a digital image can be interpreted as a signal. In this context, the aim of an age approximation technique is to detect this weak signal. In image steganalysis, the main objective is to detect whether there is a secret message (signal) hidden in a cover file (\textit{e.g., in a digital image}). The hidden (secret signal) is also a very weak signal, so that it is hard to detect for image steganalysis techniques. For this reason, the methods available in the field of image steganalysis may also be suitable for detecting existing age traces that are hidden in a digital image. The `Steganalysis Residual Network (SRNet)’ is a recent approach in the field of image steganalysis. This network is based on the residual learning principle \cite{he16a} and was introduced by Boroumand et al. in \cite{boroumand18a}. The SRNet was used in \cite{Joechl21b}, where we studied the age traces exploited by a \ac{CNN} and is also used for this work.

\begin{figure}
	\centering
	\includegraphics[width=0.98\linewidth]{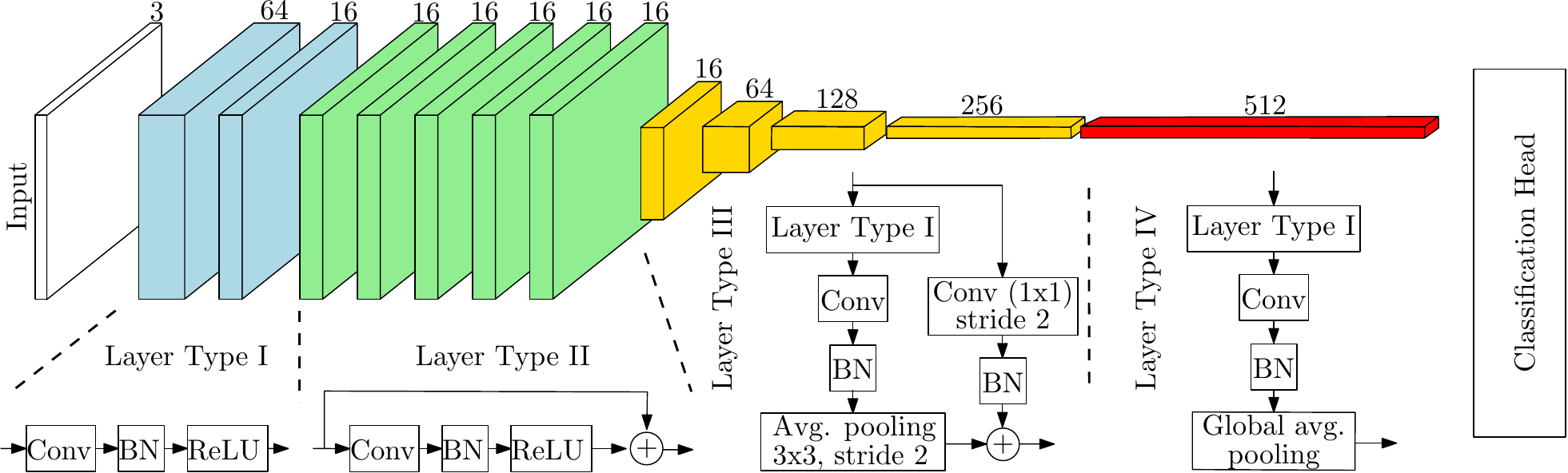}
	\caption{Overview of the SRNet \cite{boroumand18a}.}
	\label{fig:SRN}
\end{figure}

An overview of the SRNet architecture is depicted in Fig. \ref{fig:SRN}. In principle, the SRNet comprises four different layer types. The first layer type is a conventional convolution block, consisting of a convolutional layer with a $3 \times 3$ kernel size followed by a \ac{BN} and a \ac{ReLU}. The second layer type is an unpooled residual learning block. Pooling in the form of a $3 \times 3$ convolution with stride 2 is applied in the third layer type. The fourth layer consisting of a standard convolution block (layer type 1) followed by a second convolutional layer, a \ac{BN} and a global average pooling layer. The output of the fourth layer (global average pooling) is then fed into the classification head (\textit{i.e.,} a \ac{FL}) followed by a softmax function. The final network architecture is a sequence of two type 1 layers, five type 2 layers, four type 3 layers, and one type 4 layer. A key part of the SRNet are the first 7 layers (comprising layer types one and two), because there is no pooling operation involved. Since pooling acts like low-pass filtering, omitting it does not suppress the noise-like stego (or age) signal.

In \cite{Joechl21b}, several learning scenarios (based on different image patches) were used to investigate if there are additional age traces (apart from strong in-field sensor defects) learned and if these learned features are positionally invariant. The `five-crop-fusion’ scenario was the most position dependent scenario and is also used for this work. In particular, in this scenario five different SRNets are trained on five different fixed image patches each (\textit{i.e.}, each network is always trained with image patches ($256 \times 256$) extracted from the same image location). These locations are at the top left (‘tl’), the top right (‘tr’), the bottom left (‘bl’), the bottom right (‘br’) corners and in the center of the image (‘ce’). To determine the final class prediction, the outputs of each network are fused together. Further details about the image patches and the exploited SRNet can be found in \cite{Joechl21b}.

\section{Experiments}
\label{sec:Exp}
In order to investigate whether the learned features are device dependent, we train the network on images from a particular device and then apply the learned model to images from different devices. For this purpose, we use images from 14 different devices of 2 different datasets.
\subsection{PLUS Aging Datasets}
\begin{figure}
	\begin{minipage}[b]{0.98\linewidth}
		\centering
		\centerline{\includegraphics[width=1\textwidth]{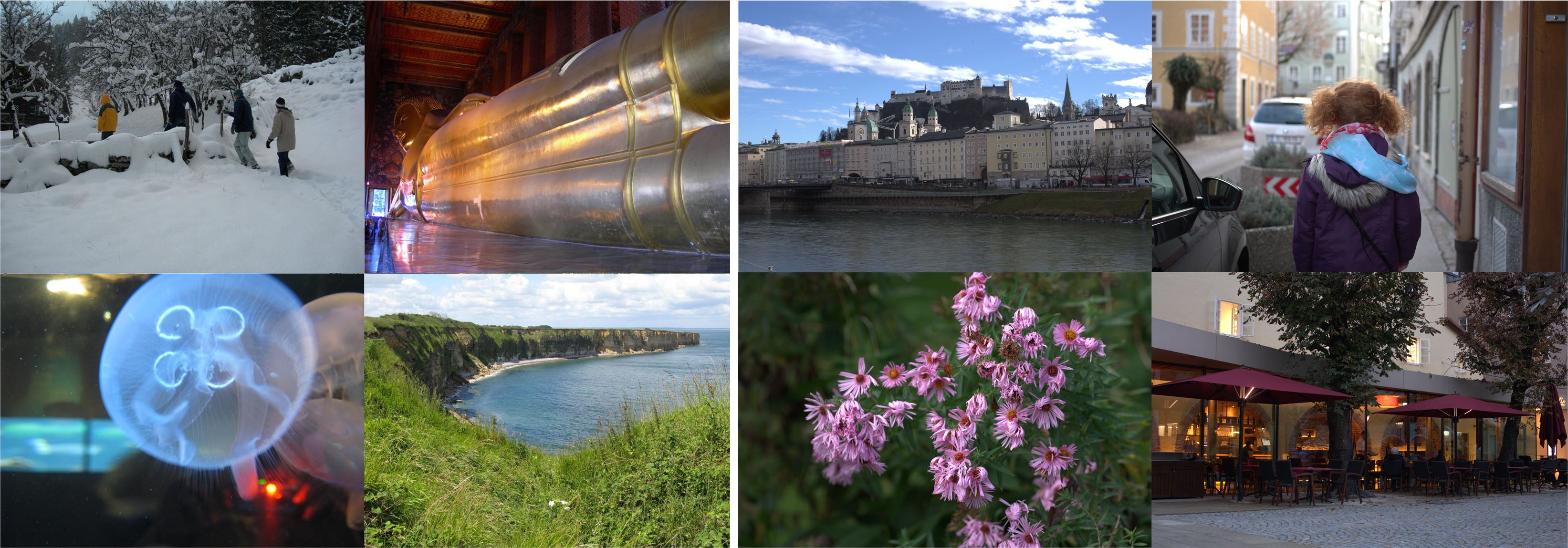}}
	\end{minipage}
	\caption{Random samples of the PLUS-nikon01 (top left), PLUS-canon01 (bottom left), PLUS-pentax01 (top right) and PLUS-pentax02 (bottom right).}
	\label{fig:PLUS-scene-samples}
\end{figure}
\begin{table}[ht!b]
    \centering
    \caption{Overview PLUS aging dataset.}
    \begin{tabular}{l l l l}
        Identifier & Make/Model & Res. [W$\times$H] & Sensor \\ \hline 
        PLUS-nikon01 & Nikon E7600 & $3072 \times 2304$ & CCD \\ 
        PLUS-canon01 & Canon PowerShotA720IS & $2592 \times 1944$ & CCD \\
        PLUS-pentax01 & Pentax K5 & $4950 \times 3284$ & CMOS\\ 
        PLUS-pentax02 & Pentax K5II & $4950 \times 3284$ & CMOS \\ 
    \end{tabular}
    \label{tab:PLUS_DS}
\end{table}
The Paris Lodron University Salzburg (PLUS) aging dataset is our own dataset where we have images from 4 different devices. For more details about the devices, see Table \ref{tab:PLUS_DS}. We consider a binary classification problem where there is a reasonable amount of time between the acquisition times of both classes. In particular, the first PLUS-nikon01 class consists of 382 images captured between October 2005 and February 2006, and the second class consists of 350 images taken between July 2019 and October 2020. The 995 PLUS-canon01 images have a time difference of about 11 years; 642 belong to the first class (June 2008 - December 2008) and 353 to the second class (January 2020 - October 2020). The PLUS-pentax01 images are divided into 316 and 343 for class one and two, respectively. The images of these two classes were taken between October 2013 and December 2013, and April 2020 and February 2021. For PLUS-pentax02, 598 images taken between October 2014 and December 2014 belong to the first class, and 227 images taken between April 2020 and February 2021 are in the second class.

All devices are used as personal devices to capture regular scene images, such as vacation scenes. In addition, both Pentax cameras (PLUS-pentax01 and PLUS-pentax02) are used for wildlife photography. Samples of the captured scenes are illustrated in Fig. \ref{fig:PLUS-scene-samples}. All images from all four devices are JPEG compressed RGB colour images. The PLUS-pentax01 and PLUS-pentax02 were originally available in raw format. JPEG compression was performed for the images of both classes at the same time and with the same settings. Additionally, we are not aware of any changes between the two classes that affect the image acquisition pipeline.

\subsection{Northumbria Temporal Image Forensics Database}
\begin{table}[ht!b]
    \centering
    \caption{Overview NTIF dataset.}
    \begin{tabular}{l l l l}
        Identifier&Make/Model&[W$\times$H]&Sensor\\ \hline 
        NTIF-canon01&Canon IXUS115HS&$4000\times3000$&CMOS\\
        NTIF-canon02&Canon IXUS115HS&$4000\times3000$&CMOS\\
        NTIF-fujifilm01&Fujifilm S2950&$4288\times3216$&CCD\\
        NTIF-fujifilm02&Fujifilm S2950&$4288\times3216$&CCD\\
        NTIF-nikon01&Nikon Coolpix L330&$5152\times3864$&CCD\\
        NTIF-nikon02&Nikon Coolpix L330&$5152\times3864$&CCD\\
        NTIF-panasonic01&Panasonic DMC TZ20&$4320\times3240$&CMOS\\
        NTIF-panasonic02&Panasonic DMC TZ20&$4320\times3240$&CMOS\\
        NTIF-samsung01&Samsung pl120&$4320\times3240$&CCD\\
        NTIF-samsung01&Samsung pl120&$4320\times3240$&CCD\\
    \end{tabular}
    \label{tab:NTIF_DS}
\end{table}
The `Northumbria Temporal Image Forensics (NTIF)’\cite{NTIFDB} database is a publicly available dataset for temporal image forensics. The NTIF dataset comprises images from 10 different cameras (of 5 different models). More details about the devices can be found in Table \ref{tab:NTIF_DS}. For each device, approximately 71 time-slots ranging over 94 weeks (between 2014 and 2016) are available. Overall, this results in 41,684 natural colour images of indoor and outdoor scenes along with 980 blue-sky scenes. Images from this database are also used in other image forensics related publications, \textit{i.e.}, \cite{alani2017, lawgaly2017, ahmed20a, ahmed21a}.
For this work, we select the first 5 time-slots as the first class and the time-slots 20 to 25 as the second class for each device. This results in around 260 images captured between October 2014 and November 2014 for the first class, and around 270 images taken between March 2015 and May 2015 for the second class. All images are JPEG compressed RGB colour images of regular scenes. A brief review of the various images suggests that the images for all 10 devices might be taken at about the same time and at about the same location.

\begin{table}[ht!b]
    \centering
    \caption{Overview average compression ratio (cr) per device.}
    \begin{tabular}{l p{1cm} | ll}
        Identifier & cr & Identifier & cr  \\ \hline 
        PLUS-nikon01 & 16.88 & NTIF-fujifilm02 & 11.15 \\
        PLUS-canon01 & 10.81 & NTIF-nikon01 & 12.23 \\
        PLUS-pentax01 & 11.91 & NTIF-nikon02 & 12.17 \\
        PLUS-pentax02 & 11.68 & NTIF-panasonic01 & 7.29 \\ 
        NTIF-canon01 & 11.6 & NTIF-panasonic02 & 7.27 \\
        NTIF-canon02 & 10.75 & NTIF-samsung01 & 13.37 \\ 
        NTIF-fujifilm01& 11.19 & NTIF-samsung02 & 13.49 \\
    \end{tabular}
    \label{tab:compression-ratios}
\end{table}

The average compression ratio of images for all 14 devices is shown in Table \ref{tab:compression-ratios}. It can be seen that very similar average compression ratios are reached for the NTIF device pairs. These almost identical average compression ratios indicate that the same compression settings were most likely used for the NTIF device pairs. Furthermore, a very similar average compression ratio at the same compression settings may also indicate that similar scenes were captured. In general, the effect of image compression on in-field sensor defect based age approximation techniques is analysed in \cite{Joechl21c}.

\subsection{Network Training and Evaluation}
For the cross-device evaluation, we train the SRNet in the `five-crop-fusion’ approach (as described in \cite{Joechl21b}) on images from a specific device. In particular, the network is trained over 80 epochs with a batch size of 4. The initial learning rate of 0.001 is decreased after 60 epochs to 0.0001. The `adamax’ \cite{kingma2017adam} is used as an optimizer with a weight decay of 0.0002. During training, the class with fewer samples is oversampled accordingly. 

A stratified sampling strategy is used for the evaluation, where 90\% of the images per class are selected as train set and 10\% are selected as test set. The evaluation is performed 10 times for each device. All train set and test set images are randomly and independently selected for each run. To assess the prediction performance of each device, the classification accuracy is computed, \textit{i.e.},
\begin{equation}
	\text{acc} = \frac{1}{N} \sum_{i}^{N} I[\hat{y} = y],
\end{equation}
where $I$ is the indicator function returning 1 only if the argument is true (\textit{i.e.}, only if the predicted class label $\hat{y}$ is equal to the true class label). The total number of test samples is denoted by $N$ . Hence, the accuracy is the ratio of correctly predicted test samples.

\section{Experimental Results}
\label{sec:Results}
\begin{figure*}[htp]
	\centering
	\begin{minipage}[b]{0.40\textwidth}
		\centering
		\includegraphics[width=0.98\linewidth]{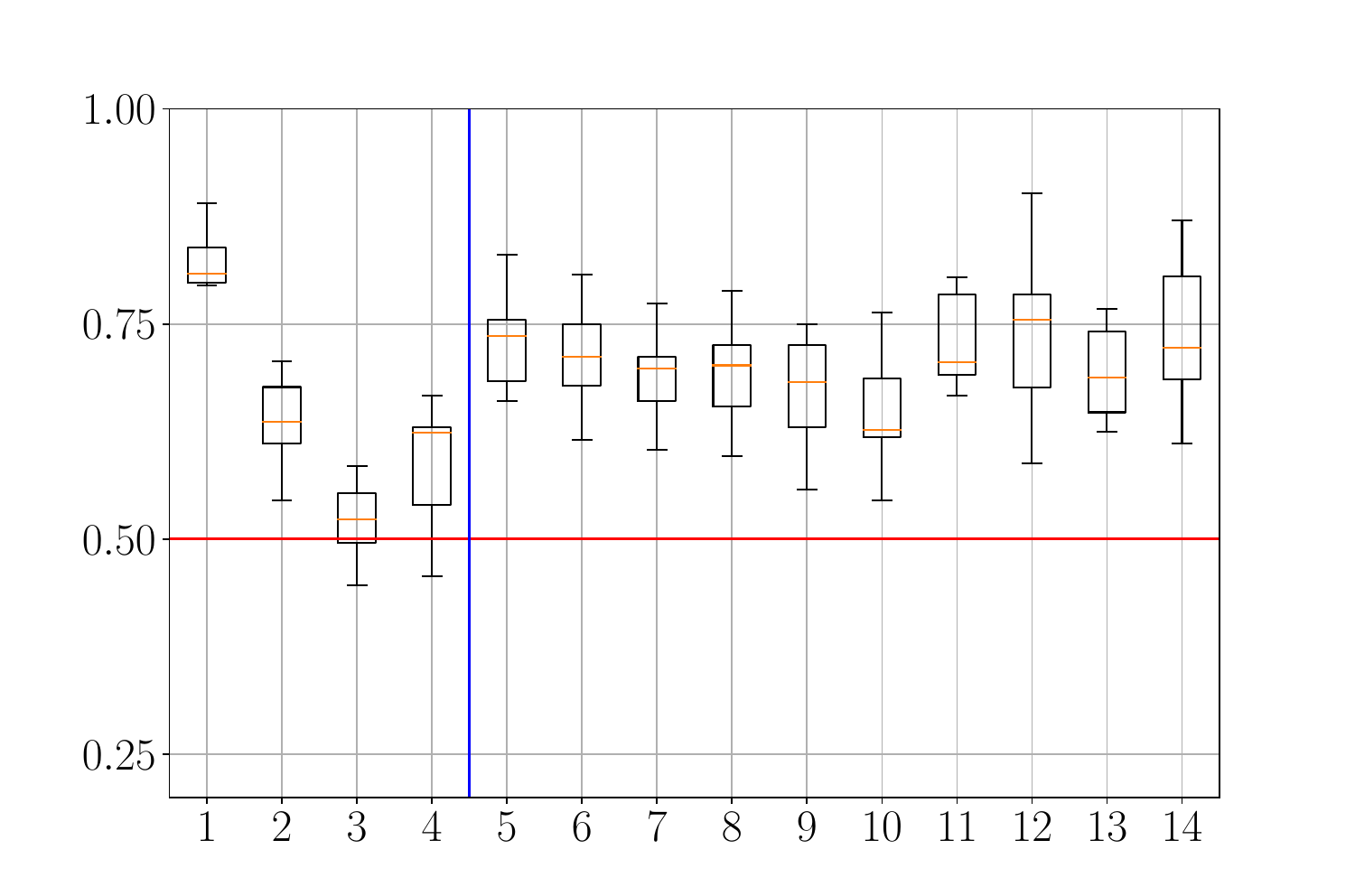}
		\centerline{\small{(a) Trained on PLUS-nikon01}}\medskip
	\end{minipage}
	\begin{minipage}[b]{0.40\textwidth}
		\centering
		\includegraphics[width=0.98\linewidth]{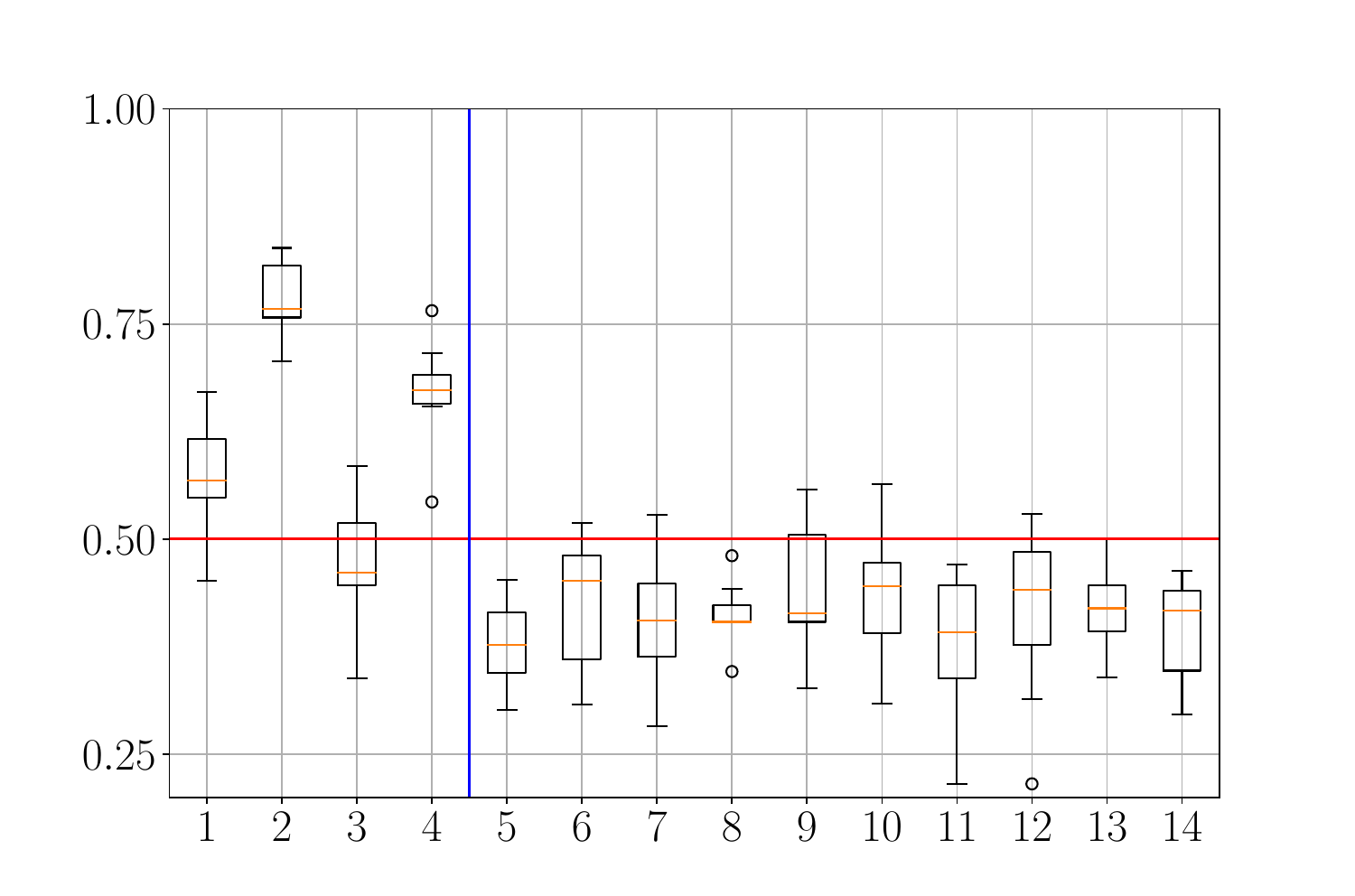}
		\centerline{\small{(b) Trained on PLUS-canon01}}\medskip
	\end{minipage}
	\begin{minipage}[b]{0.40\textwidth}
		\centering
		\includegraphics[width=0.98\linewidth]{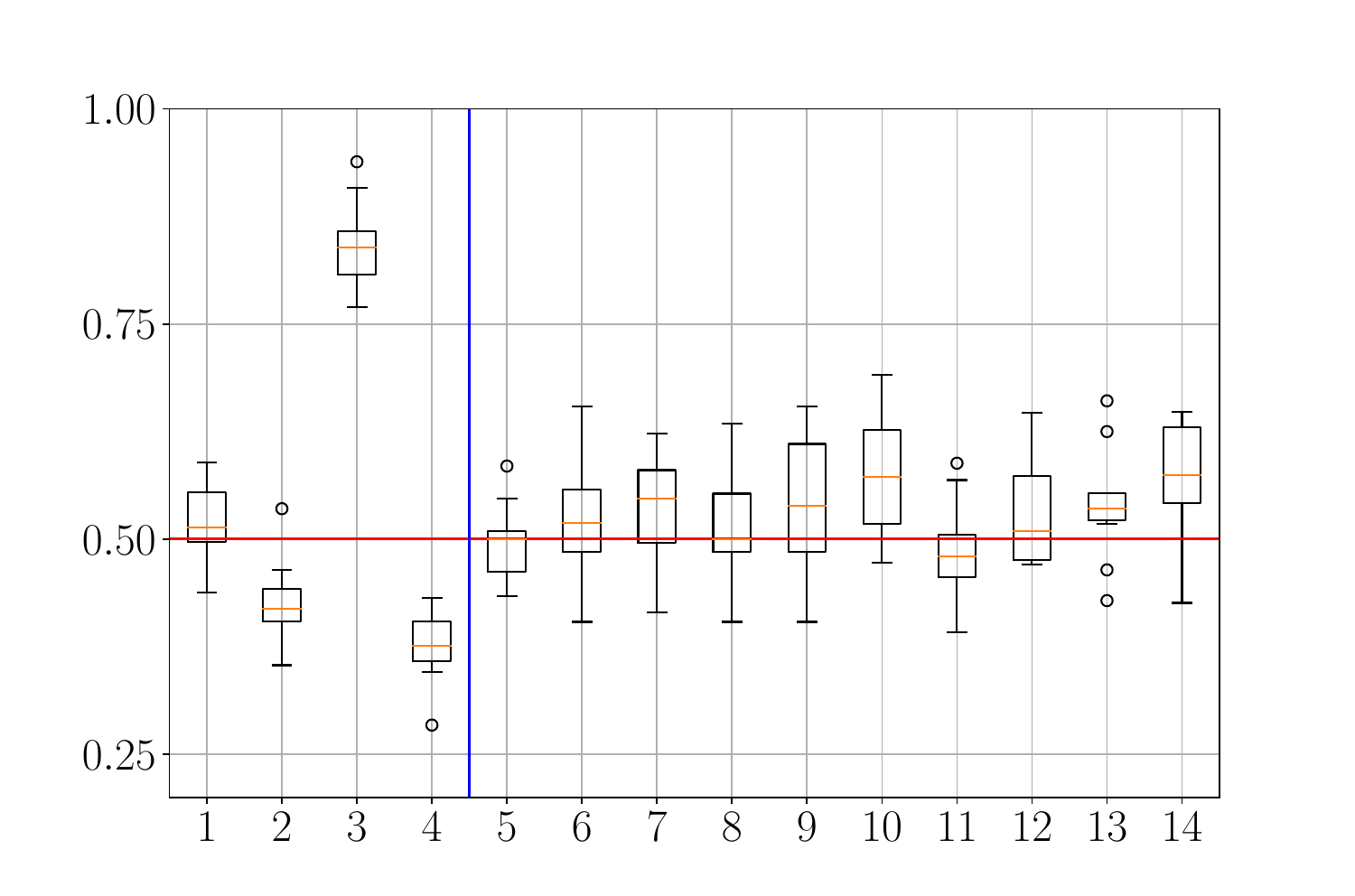}
		\centerline{\small{(c) Trained on PLUS-pentax01}}\medskip
	\end{minipage}
	\begin{minipage}[b]{0.40\textwidth}
		\centering
		\includegraphics[width=0.98\linewidth]{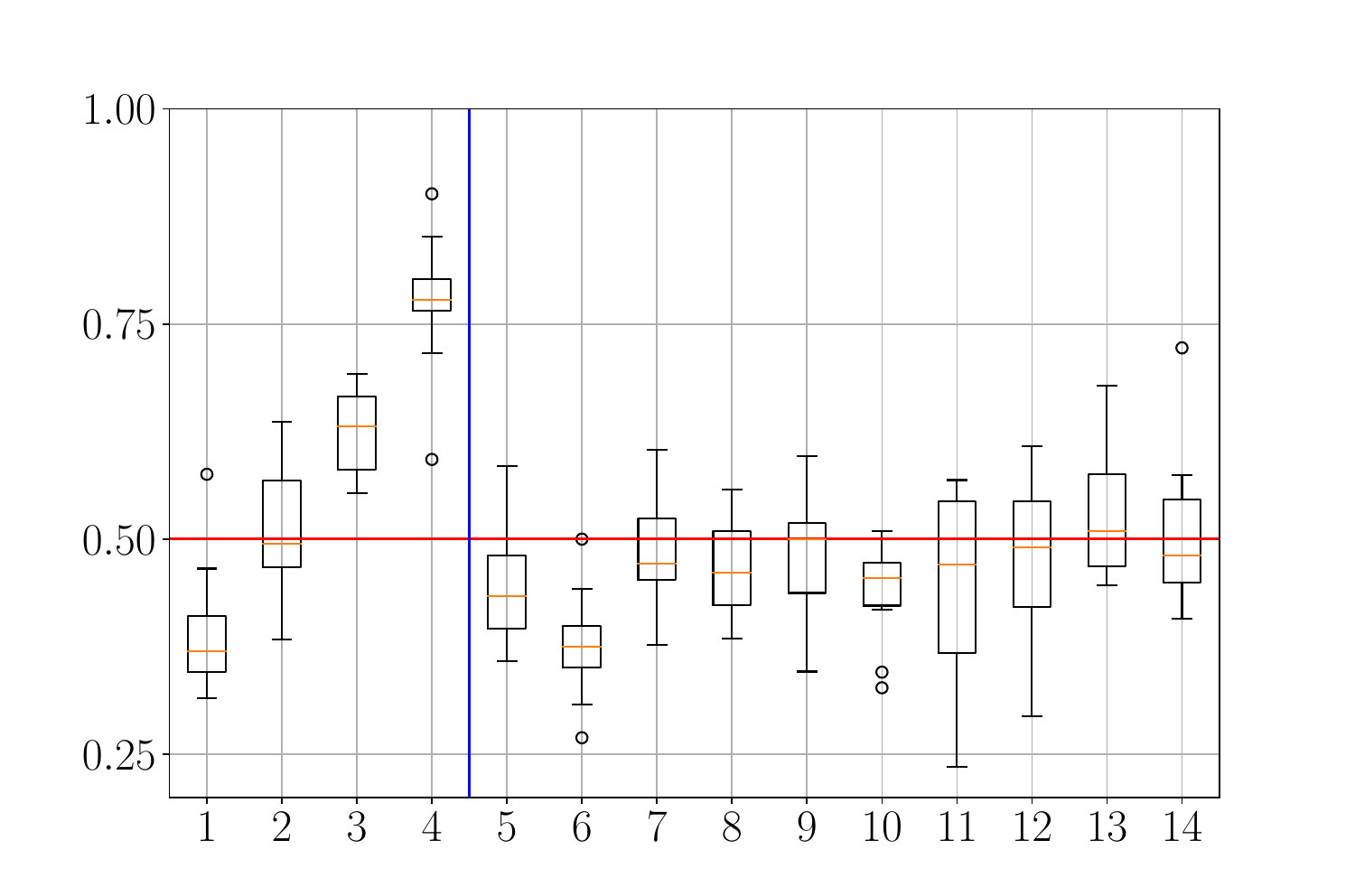}
		\centerline{\small{(d) Trained on PLUS-pentax02}}\medskip
	\end{minipage}
	\caption{PLUS - Boxplot of the resulting prediction accuracy for 10 different runs. The boxes 1-4 (left of the vertical blue line) represent the PLUS-nikon01, PLUS-canon01, PLUS-pentax01 and PLUS-pentax02, respectively. The boxes 5-14 represent the NTIF devices canon01, canon02, fujifilm01, fujifilm02, nikon01, nikon02, panasonic01, panasonic02, samsung01 and samsung02, respectively.}
	\label{fig:acc-PLUS}
\end{figure*}

\begin{figure*}[htp]
    \centering
    \begin{minipage}[b]{0.40\textwidth}
		\centering
		\includegraphics[width=0.98\linewidth]{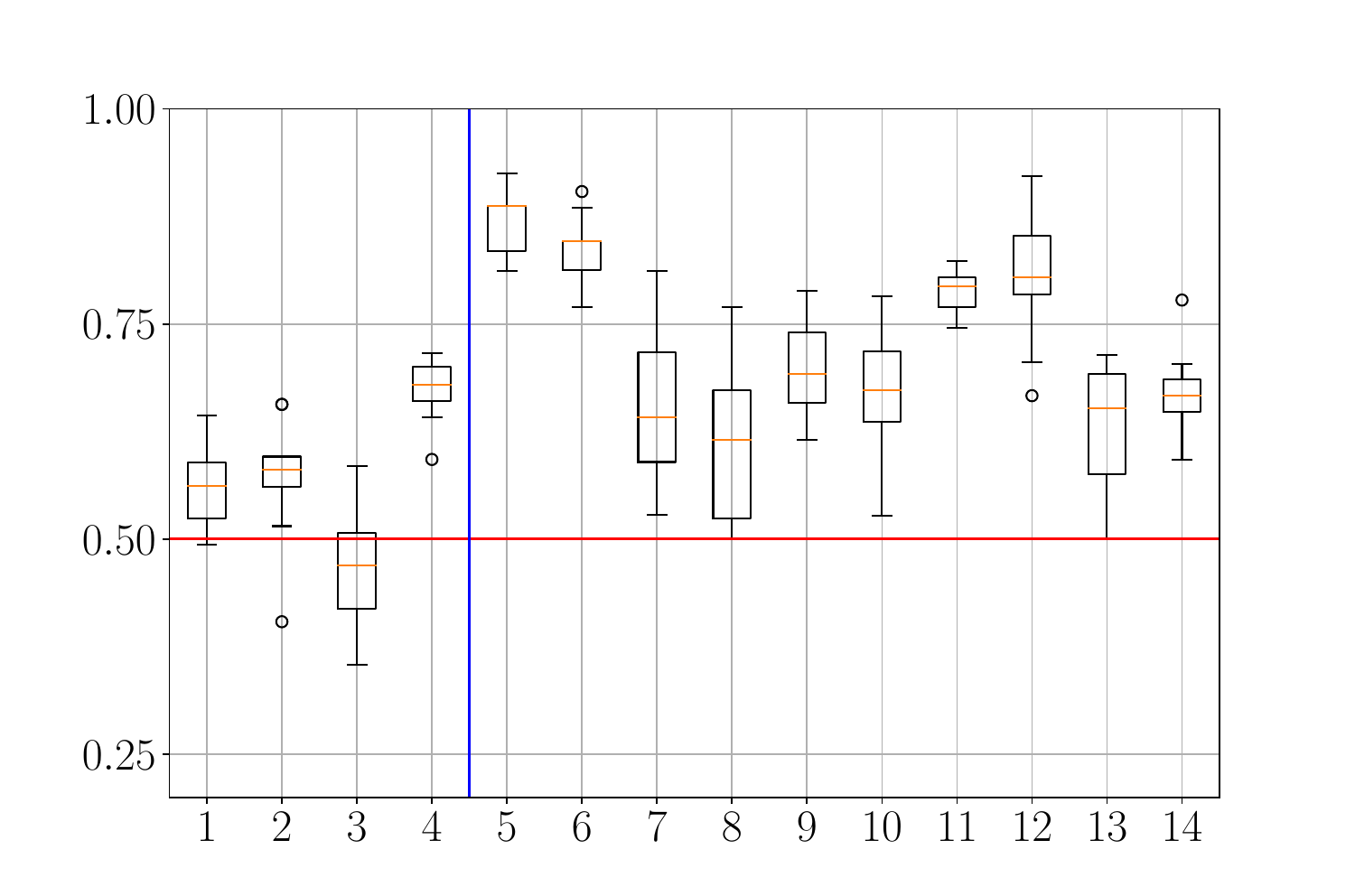}
		\centerline{\small{(a) Trained on NTIF-canon01}}\medskip
	\end{minipage}
	\begin{minipage}[b]{0.40\textwidth}
		\centering
		\includegraphics[width=0.98\linewidth]{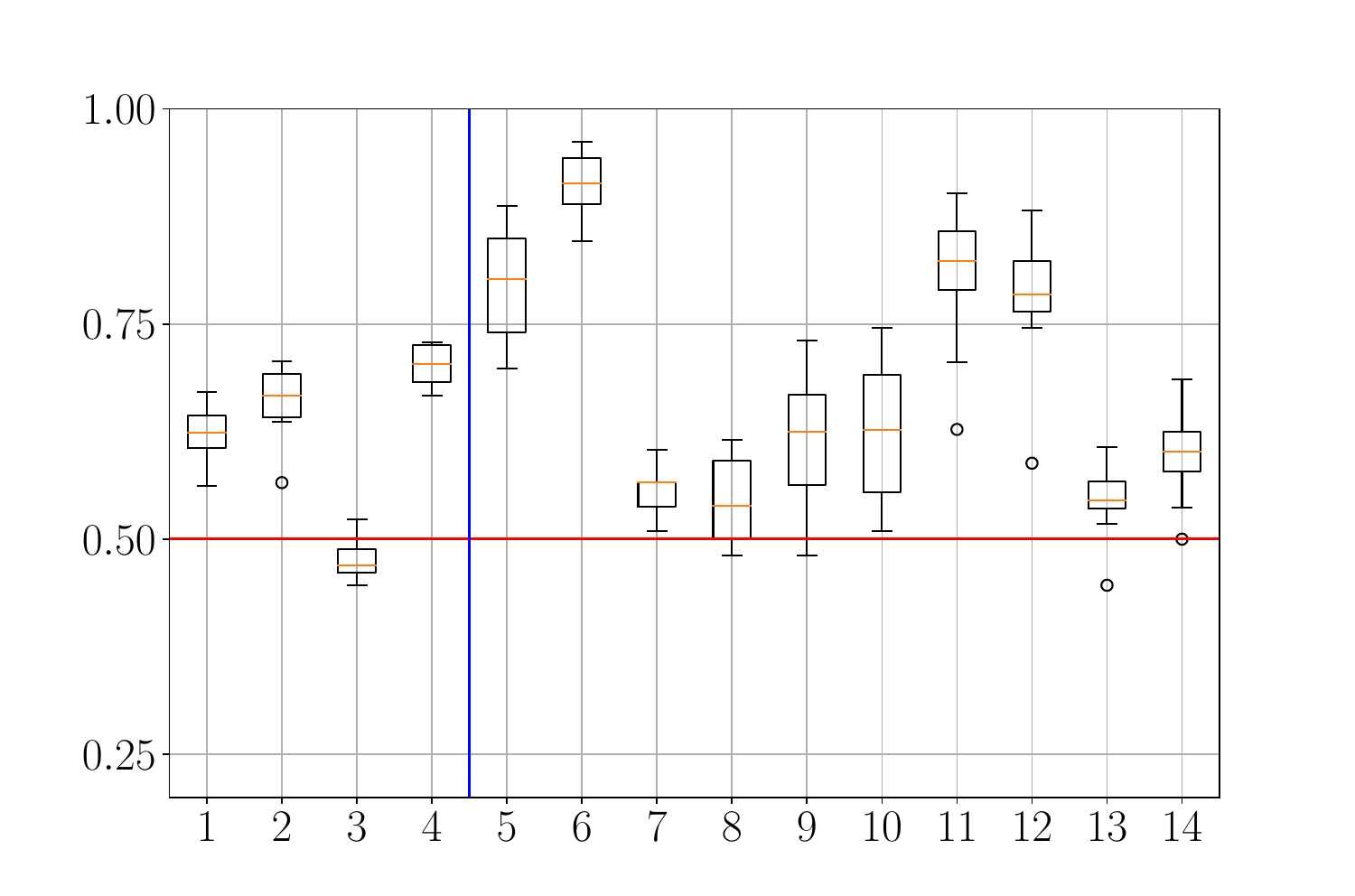}
		\centerline{\small{(b) Trained on NTIF-canon02}}\medskip
	\end{minipage}
	\begin{minipage}[b]{0.40\textwidth}
		\centering
		\includegraphics[width=0.98\linewidth]{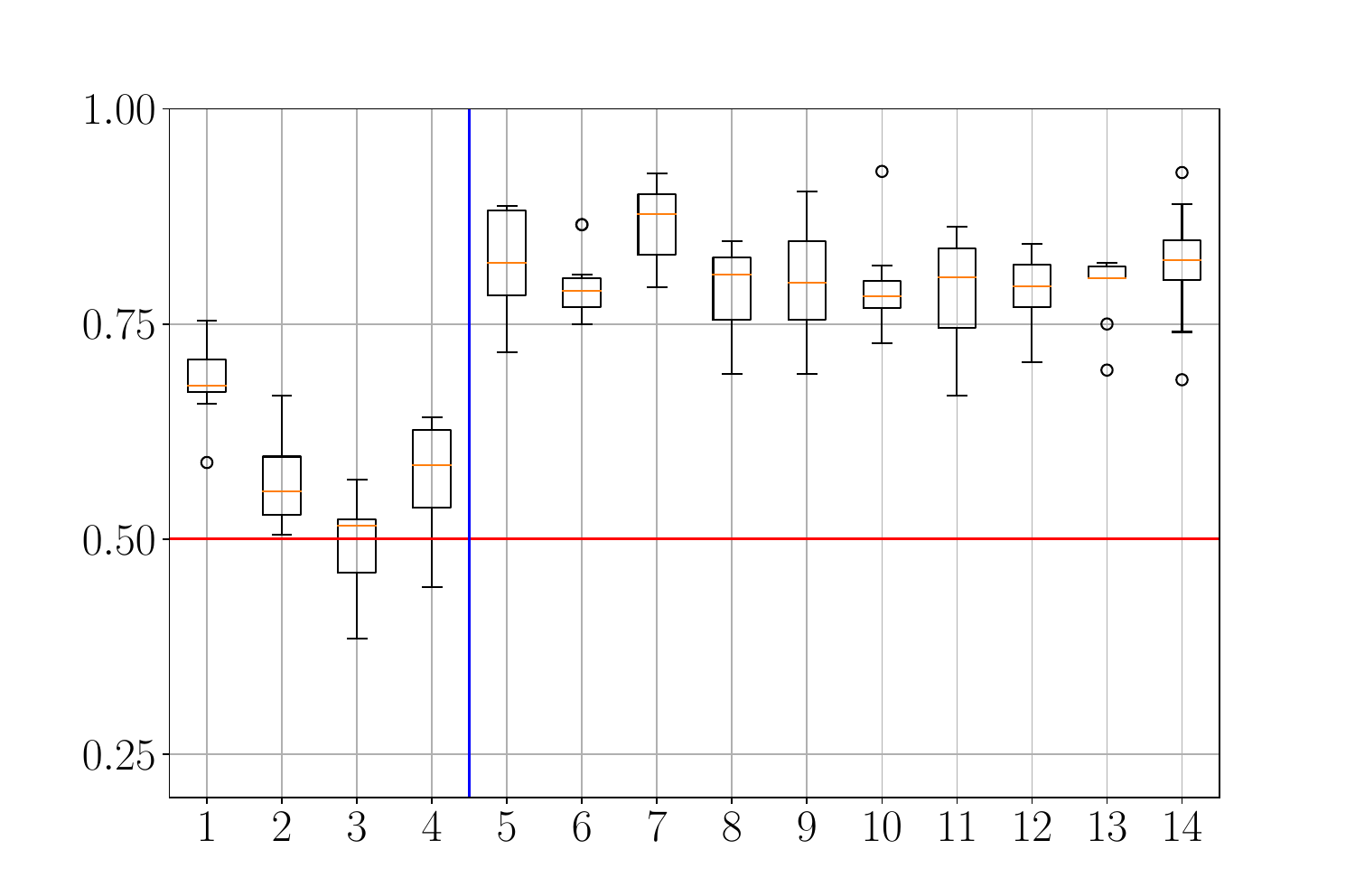}
		\centerline{\small{(c) Trained on NTIF-fujifilm01}}\medskip
	\end{minipage}
	\begin{minipage}[b]{0.40\textwidth}
		\centering
		\includegraphics[width=0.98\linewidth]{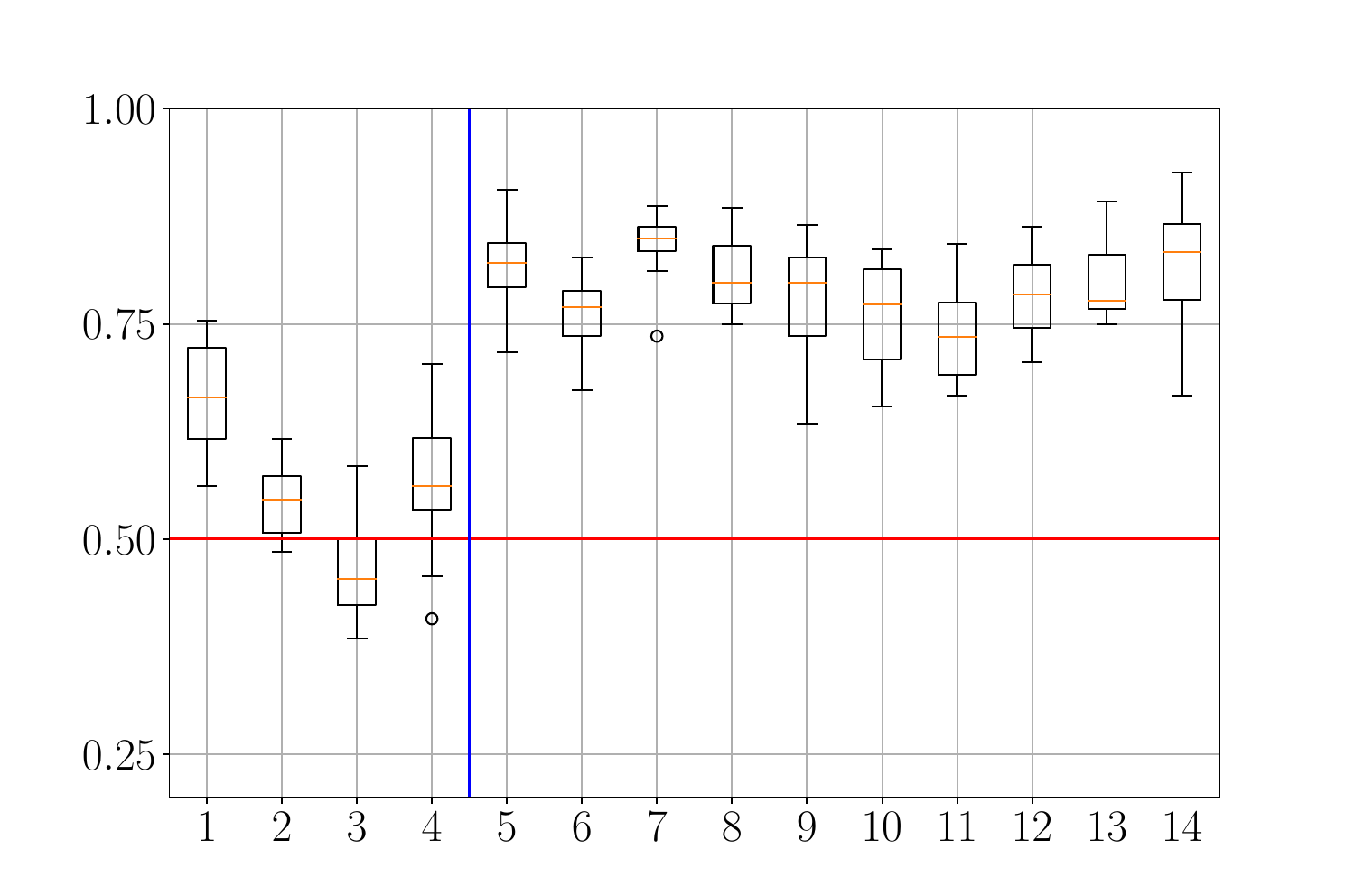}
		\centerline{\small{(d) Trained on NTIF-fujifilm02}}\medskip
	\end{minipage}
	\begin{minipage}[b]{0.40\textwidth}
		\centering
		\includegraphics[width=0.98\linewidth]{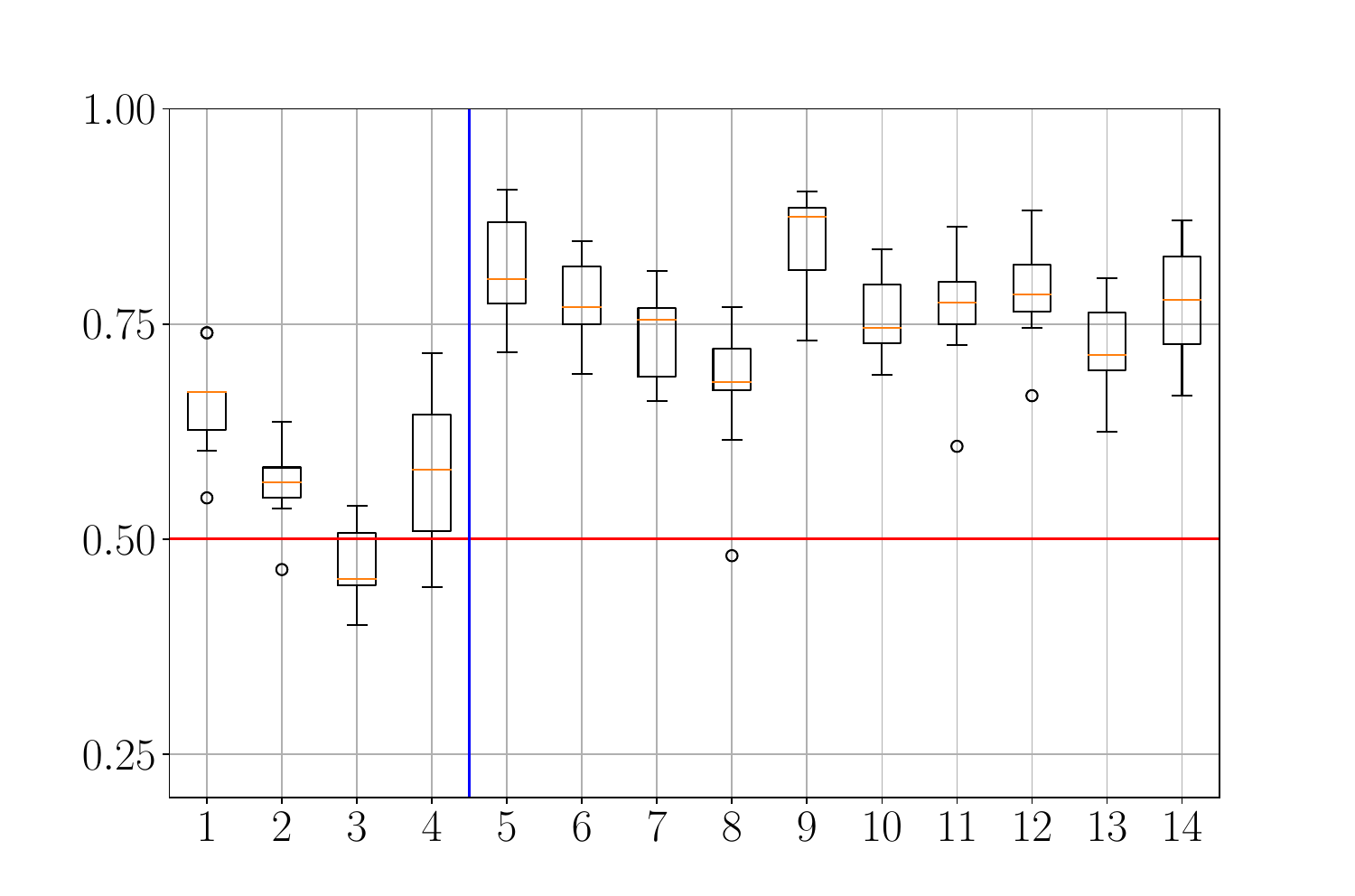}
		\centerline{\small{(e) Trained on NTIF-nikon01}}\medskip
	\end{minipage}
	\begin{minipage}[b]{0.40\textwidth}
		\centering
		\includegraphics[width=0.98\linewidth]{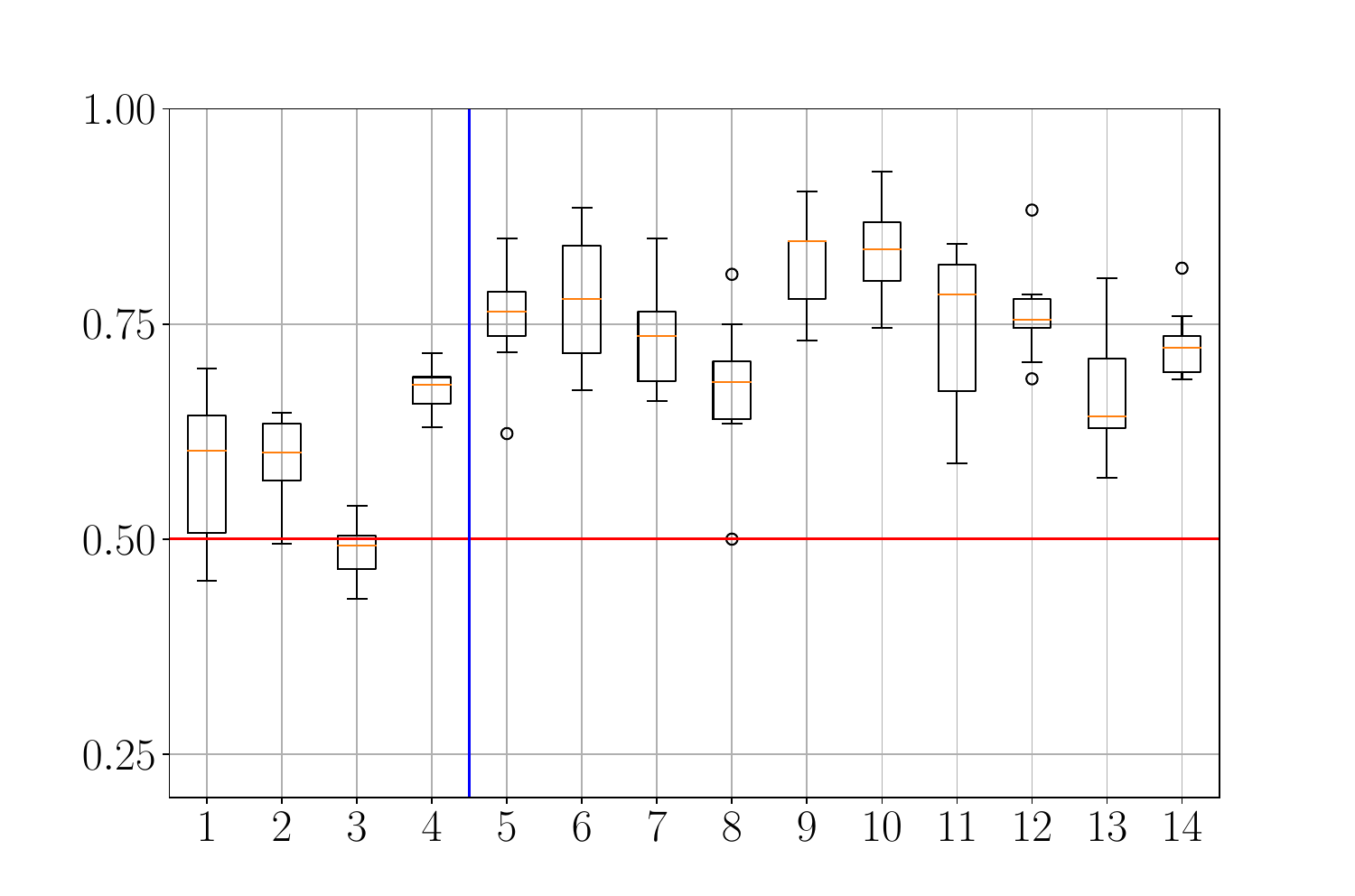}
		\centerline{\small{(f) Trained on NTIF-nikon02}}\medskip
	\end{minipage}
	\begin{minipage}[b]{0.40\textwidth}
		\centering
		\includegraphics[width=0.98\linewidth]{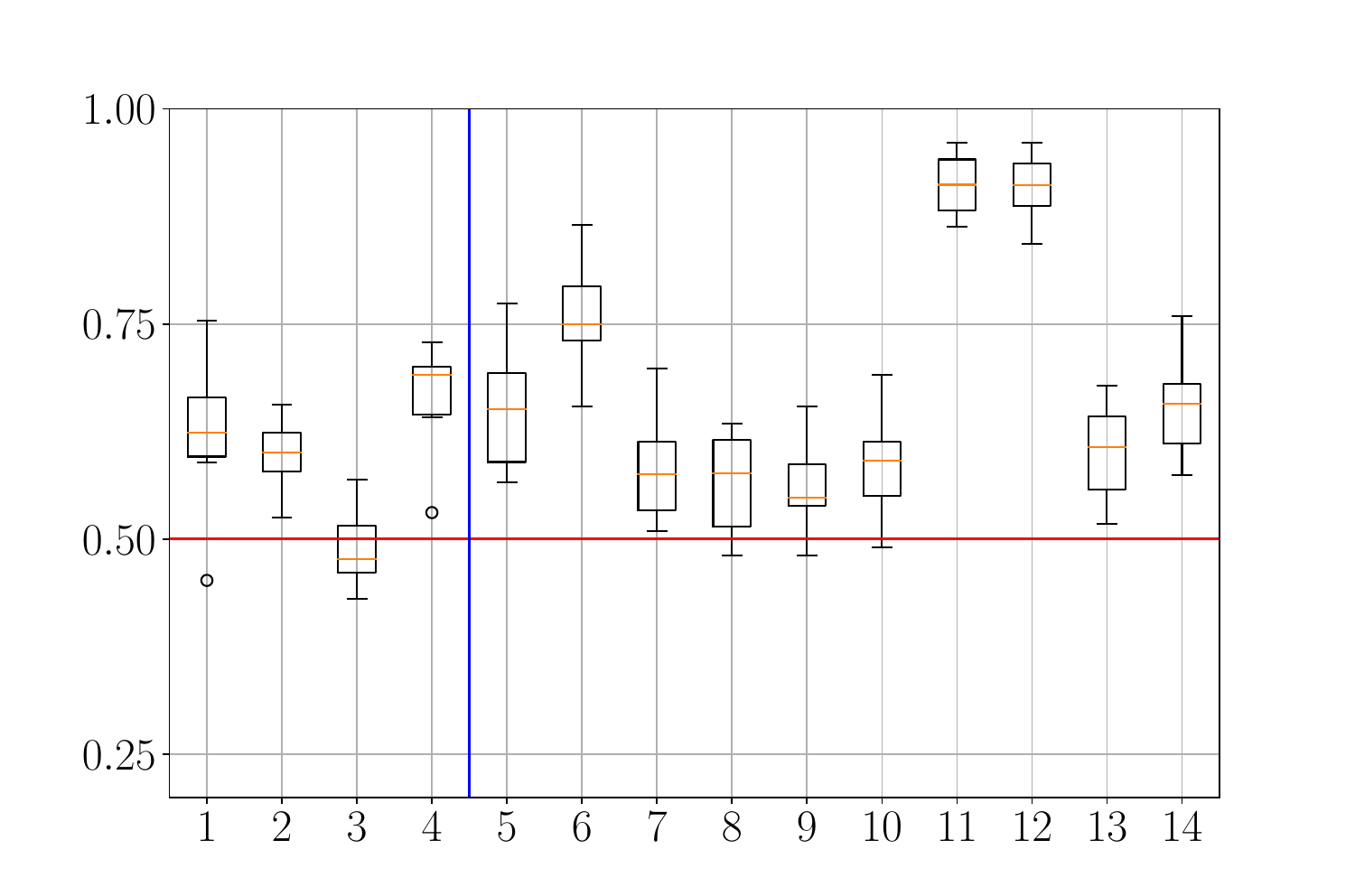}
		\centerline{\small{(g) Trained on NTIF-panasonic01}}\medskip
	\end{minipage}
	\begin{minipage}[b]{0.40\textwidth}
		\centering
		\includegraphics[width=0.98\linewidth]{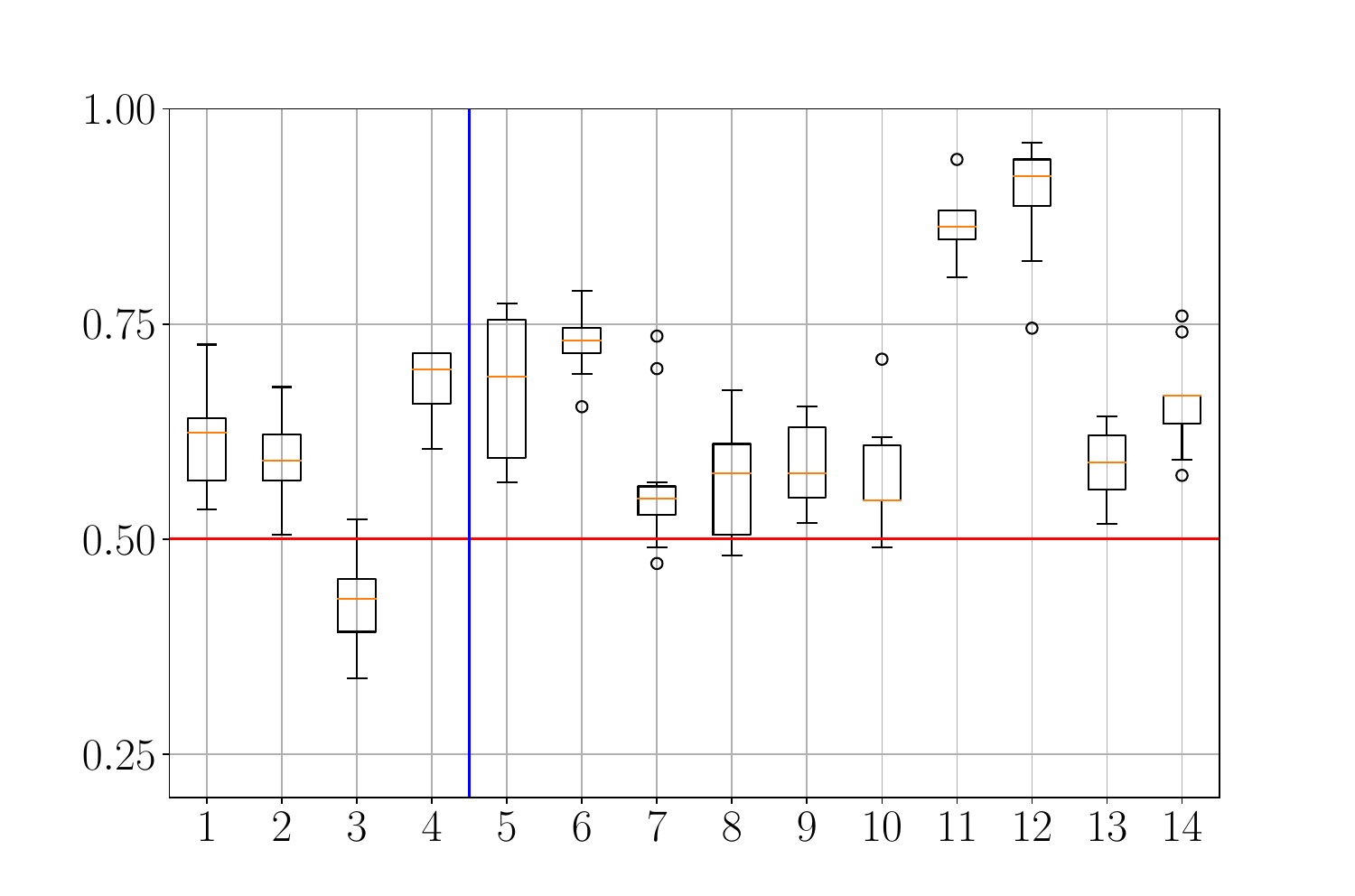}
		\centerline{\small{(h) Trained on NTIF-panasonic02}}\medskip
	\end{minipage}
	\begin{minipage}[b]{0.40\textwidth}
		\centering
		\includegraphics[width=0.98\linewidth]{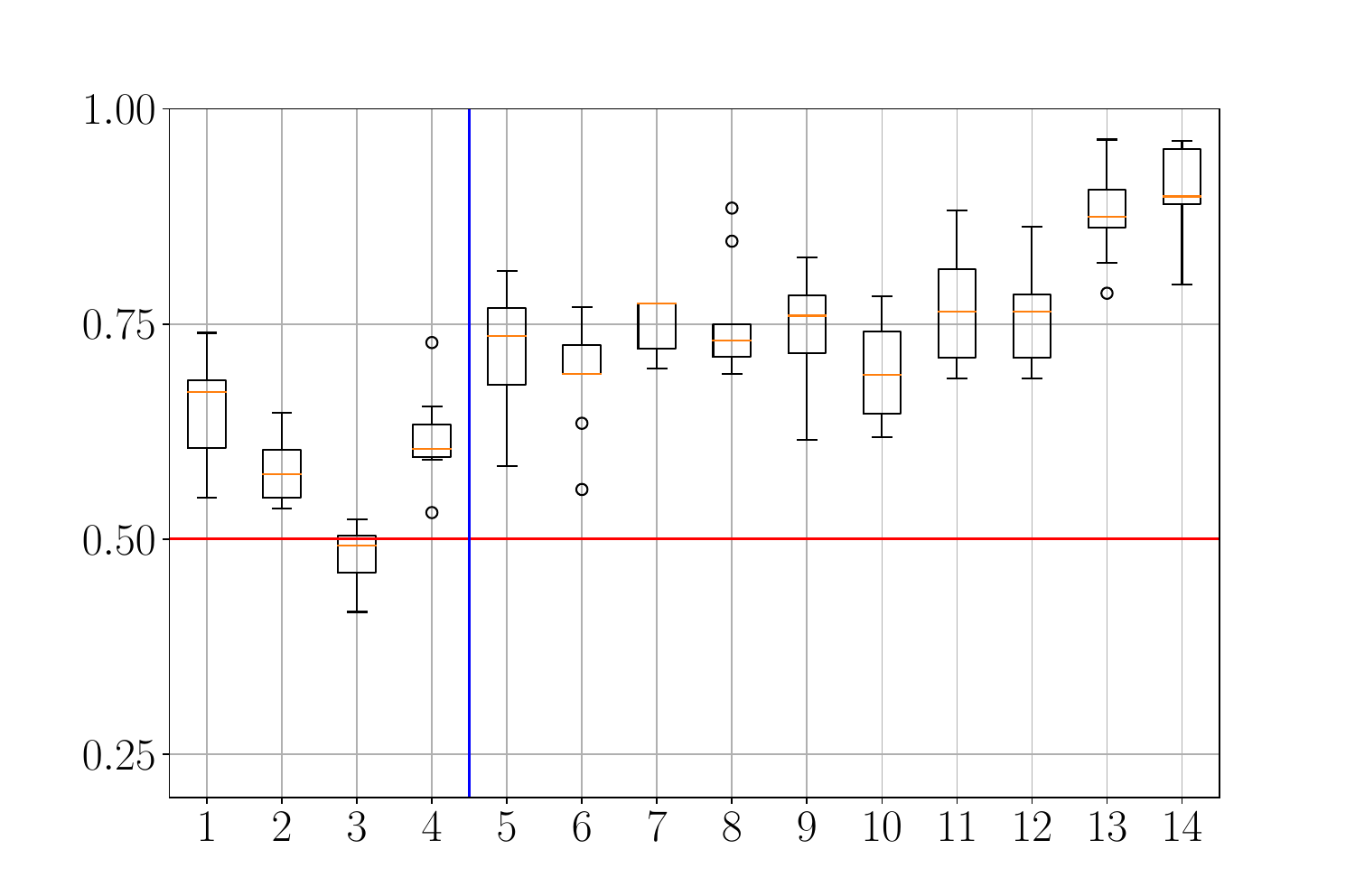}
		\centerline{\small{(i) Trained on NTIF-samsung01}}\medskip
	\end{minipage}
	\begin{minipage}[b]{0.40\textwidth}
		\centering
		\includegraphics[width=0.98\linewidth]{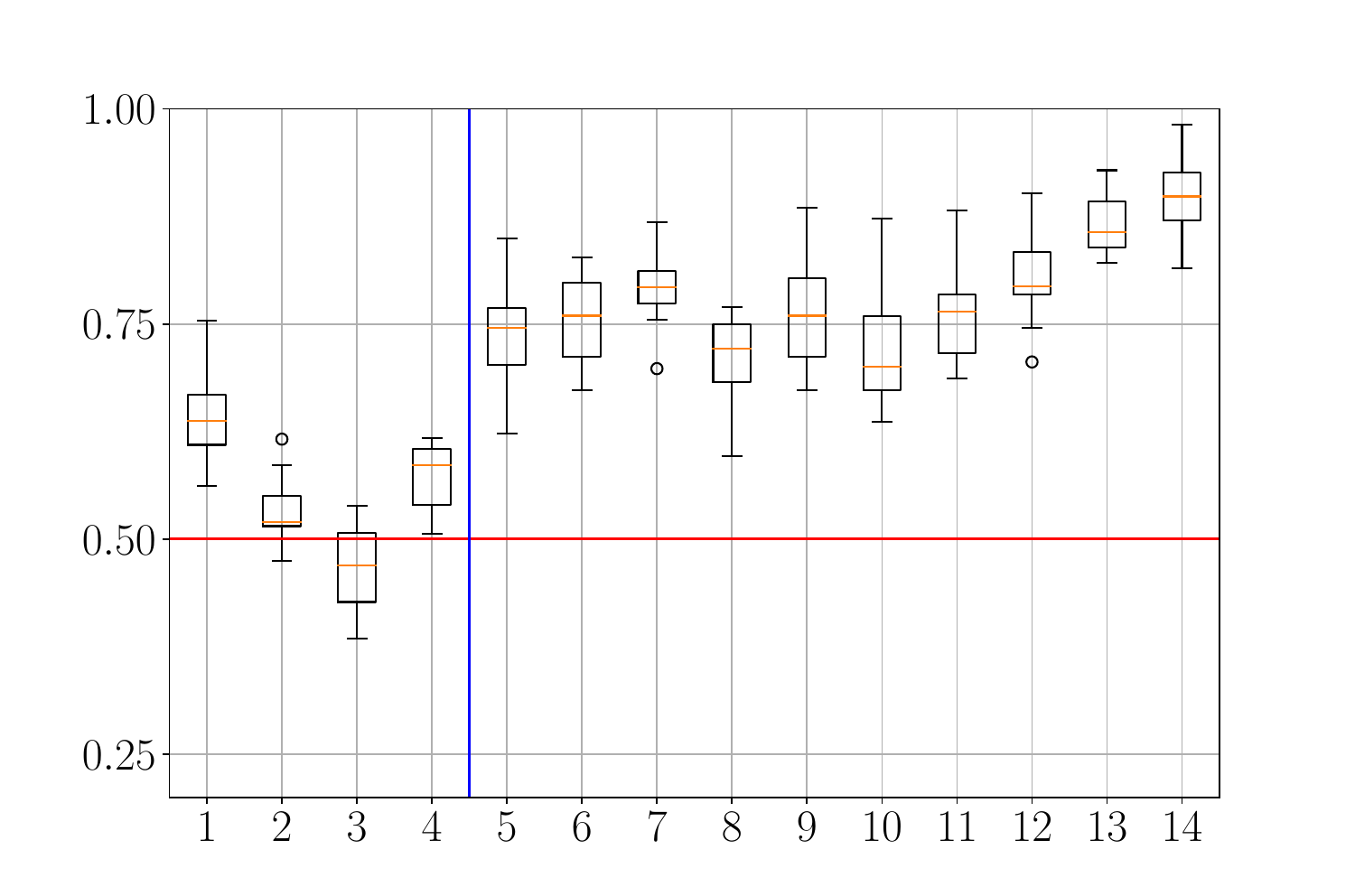}
		\centerline{\small{(j) Trained on NTIF-samsung02}}\medskip
	\end{minipage}
	\caption{NTIF - Boxplot of the resulting prediction accuracy for 10 different runs. The boxes 1-4 (left of the vertical blue line) represent the PLUS-nikon01, PLUS-canon01, PLUS-pentax01 and PLUS-pentax02, respectively. The boxes 5-14 represent the NTIF devices canon01, canon02, fujifilm01, fujifilm02, nikon01, nikon02, panasonic01, panasonic02, samsung01 and samsung02, respectively.}
	\label{fig:acc-NTIF}
\end{figure*}

In Fig. \ref{fig:acc-PLUS} and \ref{fig:acc-NTIF}, the accuracy values for 10 runs and all 14 devices are illustrated by a boxplot. The size of the box is defined by the interquartile range of the 25th and 75th percentiles of the obtained accuracy values. The median is represented by the red line, and basically all observations are within the whiskers. Outliers (\textit{i.e.}, 1.5 times the interquartile range) are marked as circles. The shown box plots in Fig. \ref{fig:acc-PLUS} and \ref{fig:acc-NTIF} are generated with the standard `matplotlib’ python implementation.

The accuracy values obtained when the model is trained on images from the PLUS Aging dataset are illustrated in Fig. \ref{fig:acc-PLUS} (a) - (d). It can be seen that, for all four different PLUS devices, the best results are achieved when the model is applied to test images of the same device (for which the model was trained on). This is reasonable since aging is typically device specific. For the PLUS-nikon01 (Fig. \ref{fig:acc-PLUS} (a)), a relatively high accuracy is achieved for 11 out of 13 devices (excluding the device on which the model is trained). This indicates that some of the learned features are device independent. However, a totally different impression is obtained when checking Fig. \ref{fig:acc-PLUS} (b) - (d). When the model is trained with images from the PLUS-canon01, PLUS-pentax01 and PLUS-pentax02, the cross-device accuracy is significantly worse than compared to the PLUS-nikon01. In contrast to the PLUS-nikon01 results, these results would indicate that the learned features are completely device dependent.

For all PLUS devices evaluated (Fig. \ref{fig:acc-PLUS} (a) - (d)) it is noticeable that the results are very similar across all NTIF devices (to the right of the blue vertical line). The results of the intra-PLUS devices, on the other hand, vary to a greater extent. Based on this observation, one might suggest that the NTIF devices share some common `age’ properties.

All results obtained when the model is trained on images from NTIF devices are illustrated in Fig. \ref{fig:acc-NTIF}. As can be seen in Fig. \ref{fig:acc-NTIF} (d) and (i), the best accuracy is not achieved when the model is applied to test images from the device on which it was trained. Instead, the best accuracy is achieved when the model is applied to test images from another device (\textit{i.e.}, trained on images from NTIF-fujifilm02 and NTIF-samsung01, while achieving best accuracy for images from NTIF-fujifilm01 and NTIF-samsung02, respectively). Looking at the results of NTIF-panasonic01 (Fig. \ref{fig:acc-NTIF} (g)) shows that the accuracy values reached when applying the model to NTIF-panasonic02 images are very similar to the results obtained when applying it to the own test images. Achieving a higher or similar accuracy on images from another device would indicate that the features learned are completely device independent (at least for the same device model). Furthermore, learning features that are completely device-independent would imply that the learned features are also fully position invariant. This contradicts the results reported in \cite{Joechl21b}, in which it is suggested that the learned features are mostly not position invariant.

When the model is trained on 3 out 5 model pairs from the NTIF dataset (\textit{i.e.}, NTIF-fujifilm Fig. \ref{fig:acc-NTIF} (c) \& (d), NTIF-nikon Fig. \ref{fig:acc-NTIF} (e) \& (f) and NTIF-samsung Fig. \ref{fig:acc-NTIF} (i) \& (j)) the accuracy reached on all other NTIF devices is considerably better than on PLUS devices. In addition, the results obtained for these devices are very similar across all NTIF imagers. Again, this would indicate that NTIF devices share some common `age’ properties that result in features which are more independent of other NTIF devices than of PLUS devices.

When considering the NTIF-canon (Fig. \ref{fig:acc-NTIF} (a) \& (b)) and NTIF-panasonic (Fig. \ref{fig:acc-NTIF} (g) \& (h)) results, it is noticeable that the different device pairs behave similarly (\textit{i.e.,} within a similar classification accuracy range). On the one hand, this could indicate that the learned age features are device model specific. On the other hand, it could also be that this is because of the same or very similar image processing pipeline of the respective device models (\textit{e.g.,} because of the same compression settings as shown in Table \ref{tab:compression-ratios}).

The overall worse results are obtained when the model is applied to PLUS-pentac01 test images (\textit{i.e.}, for 12 out of 13 devices, cross device age prediction does not work). Moreover, when the model is trained on PLUS-pentax01 images (Fig. \ref{fig:acc-PLUS} (c)), hardly any device-independent features are learned. 

In summary: (i) when the model is trained on images from the PLUS-canon01, PLUS-pentax01 and PLUS-pentax02 the learned age features are basically not device independent; (ii) when the network is trained on images from the NTIF-fujifilm02, NTIF-panasonic01 and NTIF-samsung01 the learned age features are basically fully device independent (at least for the same device model); (iii) when the network is trained on the remaining devices, the learned age features are partially device independent; (iv) for 10 of 14 devices the results across the images from NTIF devices are relatively similar (\textit{i.e.,} within a similar classification accuracy range); (v) when the model is trained on NTIF images, the results on other NTIF images tend to be better than on PLUS images.

Based on these observed results, no overall trend is visible and the question if the learned `age’ features are device (in)dependent can hardly be answered. On the contrary, the results cast doubt on whether the network learns solely age related features. As described earlier, a CNN learns the features used for classification independently. This has the advantage that additional traces (besides the known ones (\textit{e.g.}, in-field sensor defects)) can be exploited for image age approximation. However, it also carries the risk of learning features that can be used to successfully discriminate between the age classes, but that are unrelated to the temporal differences between the classes (image age). 

In general, images belonging to the same age class are taken in close temporal proximity. Thus, they most likely share some common properties, such as: (i) common scene properties (\textit{e.g.}, urban or nature scenes); (ii) common weather conditions (\textit{e.g.}, cloudy or blue sky); (iii) seasonal commonalities (\textit{e.g.}, light conditions and vegetation). Such none age related differences can also be exploited by the network. One indication of this is that the results across the NTIF devices are relatively similar for 10 out of 14 evaluated devices. As already discussed it might be that more or less similar images (at the same time and location) are taken with all NTIF devices for each age class (\textit{i.e.}, the captured scenes of the first class are similar across all NTIF images and, analogously, the scenes within the second class are similar as well). This could explain the generally better device independence between intra-NTIF devices. An additional indication that image content is more relevant than time difference is the observed trend that when the model is trained with images from an NTIF device, results tend to be better for images from other NTIF devices than for images from PLUS devices. However, to show such a correlation (between captured scenes and age classification accuracy), a more in-depth analysis of the captured scenes (\textit{e.g.,} to quantify scene similarities) has to be carried out.

The field of Explainable Artificial Intelligence (XAI) is focused on a better understanding and interpretation of the features learned. A recent survey of methods developed in this field is provided in \cite{linardatos21}. For example, class activation maps (CAMs) (\textit{e.g.,} proposed in \cite{chattopadhay18a,selvaraju17a,wang20a}) generate a saliency map for a given input image. The resulting saliency map highlights those image regions that influenced the class prediction. Based on these generated saliency maps, it could be assessed whether image content (independent of the embedded age signal) also influences the age class prediction.

\section{Conclusion}
In contrast to traditional image age approximation techniques (\textit{e.g.} methods based on in-field sensor defects), a \ac{CNN} based approach independently learns the (age) features used. This raises the possibility that various age features are learned, including those that have not yet been explored (as suggested in \cite{Joechl21b}). Such rich age features (\textit{e.g.,} as compared to features based only on the presence of in-field sensor defects) have the potential to considerably improve the classification accuracy and temporal resolution. However, a drawback of these independently learned features is that the origin and properties are unknown. For this reason, we started to investigate these traces in \cite{Joechl21b}. One finding was that the presence of strong in-field sensor defects in the image patches used is irrelevant to improving classification accuracy, and that the learned age features are likely not positionally invariant.

In this work, we investigated these features further. The main objective was to evaluate if these learned age features are entirely device dependent. This was done by training the model on images from a specific device and then apply the model on images from different devices. In total, 14 different devices were evaluated. However, based on the results obtained no overall trend was observable. For this reason, the question if the learned features are device (in)dependent can hardly be answered. 

In contrast, the reported results suggest that the network does not use only age related features to correctly predict the age class. As already discussed, age traces can be interpreted as a weak signal that is hidden in a digital image. Nevertheless, this embedded age signal might not be the only difference between age classes of natural scene images. Other (not image age related) intra-class properties (\textit{e.g.}, common light conditions or scenes) can distract the network from learning age related features.

For future work, it would be interesting to apply methods proposed in the field of XAI to analyse the `age’ features learned. Furthermore, to avoid distractions by common non-age related intra-class properties, a new temporal image forensics dataset should be created that eliminates potential scene or environmental dependencies. This could be achieved by capturing standardized scenes, as was done for creating the Dresden DB\cite{gloe09a}. However, in the context of image age approximation, the same standardized scenes should be captured for each time-slot.

\bibliographystyle{splncs04}
\bibliography{bibliography}
\end{document}